\documentclass[aps,prx,twocolumn,groupedaddress,amsmath,epsfig,amssymb,eqsecnum]{revtex4-1}
\usepackage{makecell}
\usepackage{bbm}
\usepackage{mathrsfs}
\usepackage{subfigure}
\usepackage{multirow} 
\usepackage{bigstrut} %
\usepackage{bm}
\usepackage{amsfonts}
\usepackage{color}
\usepackage{graphicx}
\def\newblock{\hskip .11em plus .33em minus .07em}
     \usepackage{hyperref} 
 \hypersetup{
 colorlinks=true,
 linkcolor=blue,
 urlcolor=blue,
 }

\newcommand{\dbar}{d\hspace*{-0.08em}\bar{}\hspace*{0.1em}}
\newcommand{\LFP}{{\mathscr{L}}}

\newcommand{\Hv}{\boldsymbol H}
\newcommand{\mv}{{\boldsymbol m}}

\newcommand{\be}{\begin{equation}}
\newcommand{\ee}{\end{equation}}
\newcommand{\ba}{\begin{eqnarray}}
\newcommand{\ea}{\end{eqnarray}}

\begin{document}
\title{ Stochastic Thermodynamics of Micromagnetics}
\author{Mingnan Ding$^1$}
\author{Jun Wu$^1$}
\author{Xiangjun Xing$^{1,2,3}$}
\email{xxing@sjtu.edu.cn}
\address{$^1$Wilczek Quantum Center, School of Physics and Astronomy, Shanghai Jiao Tong University, Shanghai 200240, China \\
$^2$T.D. Lee Institute, Shanghai Jiao Tong University, Shanghai 200240, China\\
$^3$Shanghai Research Center for Quantum Sciences, Shanghai 201315, China}
\date{\today} 
    
    
\begin{abstract}
In this work, we study the stochastic thermodynamics of micro-magnetic systems.  We first formulate the stochastic dynamics of micro-magnetic systems by incorporating noises into Landau-Lifshitz (LL) equation, which describes the irreversible and deterministic dynamics of magnetic moments. The resulting stochastic Landau-Lifshitz (sLL) equation obeys detailed balance, which guarantees that, with the external field fixed, the system converges to thermodynamic equilibrium with vanishing entropy production and with non-vanishing probability current. We then discuss various thermodynamic variables both at the trajectory level and at the ensemble level, and further establish both the first and the second laws of thermodynamics. Finally, we establish the Crooks fluctuation theorem, and verify it using numerical simulations.

\end{abstract}
\maketitle 

\section{Introduction} 
\label{sec:intro}

Dynamic micromagnetics~\cite{dyn-micromagnetics-Fidler,dyn-micromagnetics-Miltat,dyn-micromagnetics-Abert,dyn-micromagnetics-Chubykalo,Mayergoyz-book-2009} is the study of time evolution of magnetic moments at sub-micron scales subjected to non-steady external field or interactions.  Due to the immense importance of magnetic technology in the modern world, dynamic micromagnetics has received substantial attention  from both engineers and physicists. Traditionally, dynamics of magnetization is described by Landau-Lifshitz equation~\cite{LL-1935,LL-80-years}, or its Landau-Lifshitz-Gilbert variant~\cite{Gilbert1955,Gilbert2004,Lakshmanan2011}, which incorporates both precession and damping, but ignores thermal noises.  If the system is very small in size, or close to Curie temperature, thermal fluctuations must be incorporated into the dynamic equation.  This was first done by Brown in 1963~\cite{Brown1963} in a non-rigorous way.  Much later Aron et. al.~\cite{Aron2014} showed that in order for the magnitude of the magnetization to be conserved, the stochastic product in the equation must be interpreted in Stratonovich's way. 

More importantly, in view of the huge progresses achieved recently in stochastic thermodynamics~\cite{Jarzynski-review,Jarzynski-2004-HMF,Seifert-review,Evans-Searles,Hatano-Sasa-2001}, which is tailor-designed for non-equilibrium small systems, it is natural to ask whether such a formalism can be used to understand non-equilibrium thermodynamics of micromagnetics.  Except for a few isolated works~\cite{Bando2015,Utsumi2015}, however, the stochastic thermodynamics of micromagnetics has largely remained unexplored.  
This is probably related to two unusual features of Landau-Lifshitz dynamics that do not fit well with the standard formalism of stochastic thermodynamics.  Firstly, both magnetization and magnetic field are odd under time-reversal symmetry.  Secondly, the totality of magnetization forms a curved state space, which results in the multiplicative nature of the noises in the corresponding stochastic dynamics.  

In this work we take the first step to fill this gap. Firstly, we formulate the stochastic dynamics of micromagnetics in the presence of thermal noises using the recently developed covariant formalism of Ito-Langevin dynamics~\cite{covariant-Langevin-2020}.  The resulting {\em stochastic Landau-Lifshitz equation} obeys detailed balance and hence describes thermal fluctuations of a magnetic moment.  The probability current can be decomposed into a reversible part and an irreversible part. The reversible current is nonzero in the equilibrium state. Secondly, we develop a theory of stochastic thermodynamics based using the general theory developed in Ref.~\cite{covariant-sto-therm-1}.  Thermodynamic variables such as heat, work, and entropy productions are defined both at the trajectory level and at the ensemble level. The entropy production is shown to be non-negative and vanishes identically only at the equilibrium state.  This establishes the thermodynamic consistency of our theory.   Finally we also establish  Crooks fluctuation theorem for the integrated work and verify it  using numerical simulation. 

The rest of this paper is divided into three parts. In Sec.~\ref{sec:LLG} we derive the stochastic Landau-Lifshitz equation.  In Sec.~\ref{sec:ST} we develop stochastic thermodynamics for sLL equation.  Finally in Sec.~\ref{sec:conclusion} we draw concluding remarks and project future directions. 

\section{The Stochastic Landau-Lifshitz equation}
\label{sec:LLG}


\subsection{The Landau-Lifshitz equation}

\begin{figure}[t!]
    \centering
    \includegraphics[width=2.3in]{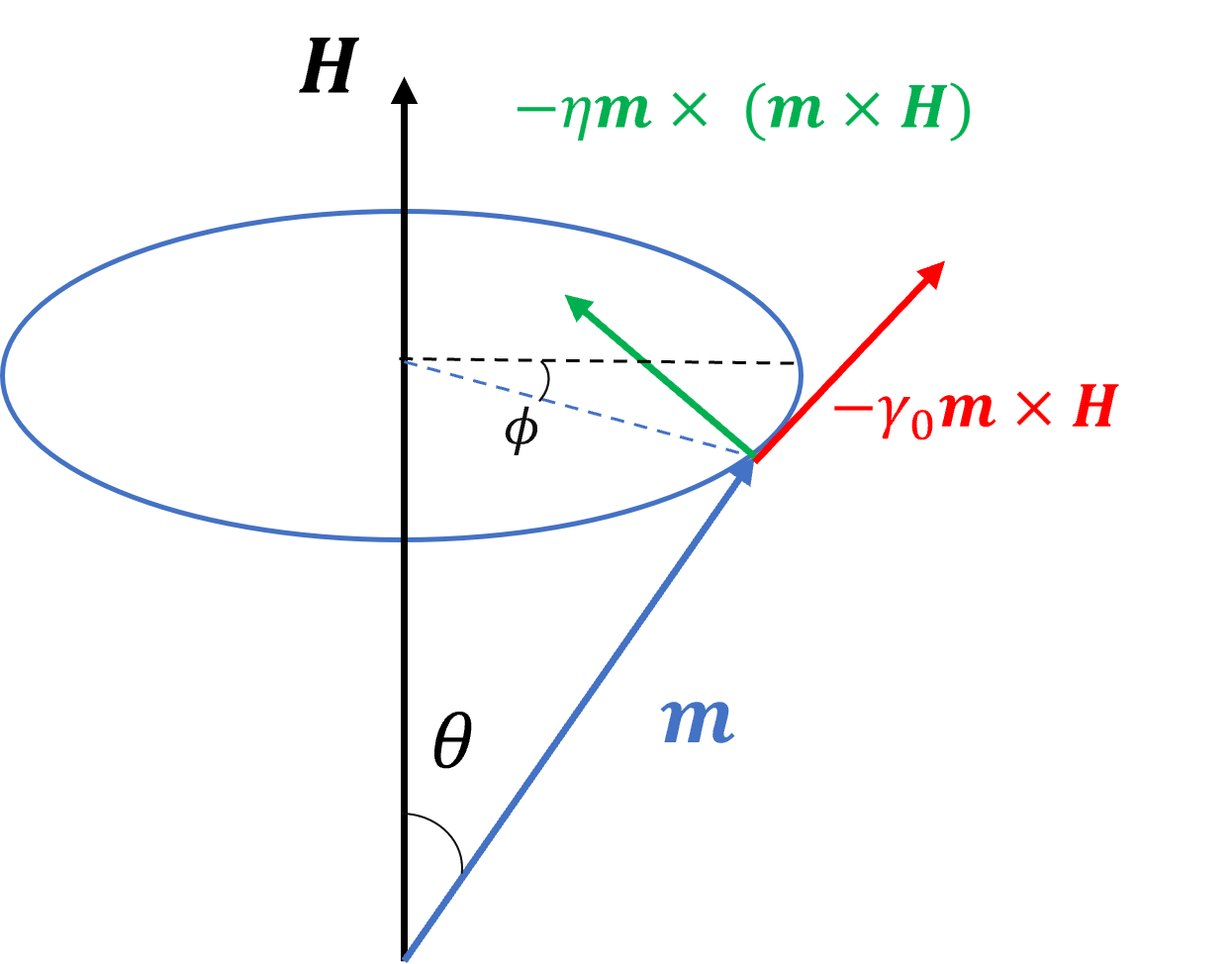}
    \caption{Schematics of Landau-Lifshitz-Gilbert dynamics. The magnitude of $\mv$ is constant.  Red arrow: $d\mv$ due to spin precession.  Green arrow: $d\mv$ due to damping. 
    }
    \label{fig:schematicsLLG}
\end{figure}

Let $\mv = (m^1, m^2,m^3)$ be the three dimensional vector of magnetization, with Cartesian components $m^1, m^2, m^3$, and $\Hv =(H_1, H_2, H_3)$ the three dimensional vector of magnetic field with Cartesian components $H_1, H_2, H_3$. 
We start from the deterministic Laudau-Lifshitz  equation which describes the dissipative dynamics of magnetic moment $\mv$ inside an effective magnetic field $\Hv$:
\be
\frac{d \mv}{dt} =  - \gamma_0 \mv \times \Hv
- \eta \, \mv \times (\mv \times \Hv),
\label{LLG-0}
\ee
where $\gamma_0$ is known as the {\em gyromagnetic ratio}, and $\eta$ shall be called the {\em damping coefficient}.  (Note that our notations are slightly different from those in previous works on LL equation.)  As illustrated in Fig.~\ref{fig:schematicsLLG}, the first term in RHS is responsible for the conservative and reversible precession of the magnetic moment around the magnetic field $\Hv$, whereas the second term describes the dissipative and irreversible damping.  In the absence of damping, the magnetic moment precess around the magnetic field with fixed angle $\theta$.  In the presence of damping, the angle $\theta$ gradually decreases, until $\mv$ becomes parallel to $\Hv$. 

Using Eq.~(\ref{LLG-0}), it is easy to verify that
\be
\frac{d \mv^2}{dt} = 2 \mv \cdot \frac{d \mv}{dt} = 0.
\ee
Hence the magnitude of $\mv$ is conserved by Eq.~(\ref{LLG-0}).  

The Laudau-Lifshitz equation (\ref{LLG-0}) may be rewritten into the  {\em Landau-Lifshitz-Gilbert} (LLG) equation:
\ba
\frac{d \mv}{dt} =  - \gamma_1 \mv \times \left(  \Hv
- \frac{\eta_1}{m} \frac{d \mv}{dt} \right),
\ea
where  $m = |\mv|$ is the magnitude of $\mv$, and $\gamma_1, \eta_1$ are related to $\gamma_0, \eta$ via
\ba
\gamma_0 = \frac{\gamma_1} { 1+ \gamma_1^2 \eta_1^2 }, \quad
\eta = \frac{\gamma_1^2 \eta_1} { m(1 + \gamma_1^2 \eta_1^2) }.
\ea
Even though the LLG equation is more popularly used in the study of micro-magnetism, the LL equation (\ref{LLG-0}) is more convenient for study of stochastic thermodynamics. 



 
 \subsection{Introducing stochasticity}
 \label{sec:sLL}
 
The stochastic version of  LL equation has been studied by several groups~\cite{Brown1963,Aron2014,Bando2015}.  A crucial issue that has not been addressed in these studies is {\em detailed balance}, which is necessary in order to guarantee that system converges to a thermodynamic equilibrium state without dissipation.  In fact, this should be deemed as the first step towards a proper formulation of stochastic thermodynamics. Here we take advantage of the covariant Ito-Langevin theory  developed systematically in previous works~\cite{covariant-Langevin-2020,covariant-sto-therm-1,sto-therm-NC}.  Let us rewrite Eq.~(\ref{LLG-0}) into the following form:
 \ba
d m^i &+& L^{ij} (\partial_j U)  \, dt = 0,
 \label{dm-LpU}
\ea
where $\partial_i = \partial/\partial m^i$, whilst  $L^{ij}$, {\em the kinetic matrix}, and $U$, {\em the generalized potential}, are respectively given by
\begin{subequations}
\label{LQBU-LLG}\ba
 L^{ij}(\mv) &=& B^{ij} + Q^{ij}, \\
Q^{ij}(\mv)  &=&   T \gamma_0 \epsilon^{ijk} m_k,
  \label{LQBU-LLG-Q}\\
 B^{ij} (\mv)&=&  T \eta \left( m^2  \delta^{ij} - m^i m^j \right), 
  \label{LQBU-LLG-B}\\
 U(\mv, \Hv) &=& - \beta \mv \cdot \Hv - \beta F (\Hv),
 \label{LQBU-LLG-U}
\ea
\end{subequations}
where $F (\Hv)$ is the {\em equilibrium free energy}:
\ba
F  (\Hv) 
&=& - T \log \frac{4 \pi m\, \sinh \beta |\Hv| m}{\beta |\Hv|}. 
\label{F-lambda-def}
\ea

%
 

Equation (\ref{dm-LpU}) is deterministic limit of the following covariant Ito-Langevin equation~\cite{covariant-Langevin-2020}:
\ba
d m^i  + \left( L^{ij} \partial_j U
- \partial_j L^{ij} \right) dt
=  b^{i \alpha} d W_{\alpha}(t) ,  \quad \quad
\label{Langevin-def-m}
\ea
where the term $- \partial_j L^{ij}$ is called {\em the spurious drift}, and $d W_{\alpha}(t) $ are the standard Wiener noises, and the product $ b^{i \alpha} d W_{\alpha}(t) $ is defined in Ito's sense.  The noise amplitudes $b^{i \alpha}$ are  related to the matrix $B^{ij}$ via
\ba
b^{i \alpha} b^{j \alpha} = 2\, B^{ij} = L^{ij} + L^{ji}.
\label{b-B-L}
\ea
This relation does not uniquely determine $b^{i \alpha} $, nor the dimension 
of Wiener noises $d W_{\alpha}(t)$.  It is most convenient to choose  $d W_{\alpha}(t)$ to be 3d and choose $b^{i \alpha}$ to be:
\ba
b^{i \alpha}(\mv) &=& 
\frac{\sqrt{2 \eta T}}{m} (m^2 \delta^{i\alpha}  - m^i m^\alpha),
\label{LQBU-LLG-b}
\ea
which clearly satisfies Eq.~(\ref{b-B-L}).  Since $b^{i\alpha}$ depends on $\mv$, the noises are {\em multiplicative}.  As explained in detail in Sec. IIC of Ref.~\cite{covariant-Langevin-2020}, in the deterministic limit, both the noise term and the spurious drift should be dropped, which reducing Eq.~(\ref{Langevin-def-m}) to the LL equation (\ref{dm-LpU}). 

Using Eqs.~(\ref{LQBU-LLG}), we may verify the following relation:
\ba
\partial_j L^{ij} &=& - 2 T \eta \, m^i.
\ea
Taking advantage of this and Eqs.~(\ref{LQBU-LLG}), we may rewrite Eq.~(\ref{Langevin-def-m}) in the following vector form:
\ba
{d \mv} =&  - & \gamma_0 \mv \times \Hv \, dt
- \eta \, \mv \times (\mv \times \Hv)\, dt
\nonumber\\
&-& 2 T \eta \,\mv \,dt 
+ {\sqrt{2 \eta T}} \mv \times d{\boldsymbol W}, 
\label{Langevin-def-m-vec}
\ea
where $d{\boldsymbol W}$ is the 3d vector-valued Wiener noise, and $-2 T \eta \,\mv dt$ is the spurious drift.  The product ${\sqrt{2 \eta T}} \mv \times d{\boldsymbol W}$ is again defined in Ito's sense.   We shall call Eq.~(\ref{Langevin-def-m-vec}) the {\em stochastic Landau-Lifshitz} (sLL) equation. It is equivalent to the sLLG equation derived by Aron {\it et. al.} in Ref.~\cite{Aron2014}.  {It is important to note that the equation of Aron et. al. does not have spurious drift because they use Stratonovich's calculus.}  Our Ito-Langevin equation (\ref{Langevin-def-m-vec}) is however more convenient for the formulation of stochastic thermodynamics.

It is important to note that both the kinetic matrix $L^{ij}$ and the generalized potential $U$ can be uniquely determined by inspecting the Landau-Lifshitz equation (\ref{dm-LpU}).  Hence the sLL equation (\ref{Langevin-def-m-vec}) is uniquely determined by its deterministic limit, the LL equation  (\ref{LLG-0}).  In order words, there is a unique covariant  Langevin dynamics corresponding to the deterministic LL equation  (\ref{LLG-0}).  

Using the basic properties of Wiener noises:
\be
dW_\alpha dW_\beta = \delta_{\alpha \beta} dt,
\ee
and Ito-Langevin equation (\ref{Langevin-def-m-vec}), we easily find
\ba
d m^2 &=& d ( \mv \cdot \mv )
= 2 \mv \cdot d \mv + d\mv \cdot d \mv 
\nonumber\\
&=&- 4 T \eta \, m^2 dt + 4 T \eta \, m^2 dt  = 0. 
\ea
Hence the magnitude of $\mv$ is conserved by sLL dynamics.  Note that it would not be so if we miss the spurious drift term in Eq.~(\ref{Langevin-def-m-vec}).  

Strictly speaking, the state space of $\mv$ is a sphere with constant radius $|\mv|$.  This is a curved space, which apparently enforces us to introduce non-Cartesian coordinates.  However, it we forget the fact that $|\mv|$ is fixed, i.e., if we consider ensemble of many magnetic moments with different $|\mv|$, we may treat $\mv$ as a three dimensional vector, and use its Cartesian components $m^1, m^2, m^3$ as the basic variables.  This viewpoint results in huge simplification, and therefore will be taken in our theory.  When we calculate averages or partition functions, of course, we sill need to restrict the integrations on the sphere with constant $|\mv|$.  



The Fokker-Planck equation (FPE) corresponding to the Ito-Langevin equation (\ref{Langevin-def-m}) is 
\ba
\partial_t \, p(\mv,t) 
= \LFP  \, p(\mv,t) ,
\label{FPE-1} 
\ea
where $p(\mv,t) $ is the probability density function of $\mv$, and $\LFP  $ is  the Fokker-Planck operator~\cite{covariant-Langevin-2020}:
\be
\LFP \equiv \partial_i L^{ij} ( \partial_j + (\partial_j U ) ).  
\label{FPO-def-0}
\ee
Using Eqs.~(\ref{LQBU-LLG}), we may rewrite $\LFP  $ into:
\ba
\LFP   &=& T\eta (m^2 \nabla^2 - m^i m^j \partial_i \partial_j - 2 \mv \cdot \nabla )
\nonumber\\
&+& \eta (- m^2 \Hv \cdot \nabla + (\mv \cdot \Hv) \mv \cdot \nabla
+ 2 \mv \cdot \Hv)
\nonumber\\
&+& \gamma_0 (\mv \times \Hv) \cdot \nabla. 
\label{FPO-LLG}
\ea
where ${\boldsymbol \nabla}$ means gradient with respect to $\mv$. 

It is straightforward to verify that the FPE (\ref{FPE-1}) admits a steady state solution:
\ba
p^{\rm eq}(\mv) =  e^{-U(\mv; \Hv)}.
\label{p-eq-U-1}
\ea
As we shall demonstrate below, the LL dynamics satisfies detailed balance.  Hence this steady state is actually a thermal equilibrium state. 


%


The FPE (\ref{FPE-1}) may be rewritten into the following form of probability conservation:
\ba
\partial_t p = \partial_k j^k, 
\label{FPE-j}
\ea
where $j^i$ is the probability current defined as:
\be
j^k = -  L^{kj} (\partial_j + \partial_j U) p + \partial_j (Q^{kj} p),
\label{current}
\ee
which can be conveniently decomposed into a {\em reversible current}  $j^i_{\rm R}$ and an {\em irreversible current} $j^i_{\rm IR}$:
\begin{subequations}
\ba
j^k_{\rm R} &=& -  Q^{kj} (\partial_j + \partial_j U) p + \partial_j (Q^{kj} p),
\label{j_R-def}
\\
j^k_{\rm IR} &=& -  B^{kj} (\partial_j + \partial_j U) p. 
\label{j_IR-def}
\ea
\label{j_R-IR-def}
\end{subequations}
In the equilibrium, only the reversible current persists:
\be
j^k_{\rm eq} = j^k_{\rm R} 
=  \partial_j (Q^{kj} e^{- U}).
\ee
Using Eq.~(\ref{LQBU-LLG}), we may rewrite it into the vector form:
\ba
{\mathbf j}_{\rm eq} = \gamma_0 \, \mv \times \Hv \, e^{-U}.
\ea
This non-vanishing equilibrium probability current is due to the reversible spin precession dynamics, and, as we shall show below, makes no contribution to dissipation.  Such a peculiarity does not show up in systems with only even variables and even control parameters. 

\subsection{FPE in spherical coordinates}
Here we rewrite the FPE (\ref{FPE-1}) in spherical coordinates and compare it with the result of Brown~\cite{Brown1963}.  The spherical coordinates of $\mv$ and $\Hv$  are respectively $\{ \tilde m_i\}  = \{ m, \theta, \phi\}$ and $\{ \tilde H_i \} = \{H, \theta_H,\phi_H \}$, which are related to their Cartesian components as
 \begin{subequations}
\ba
m^1 & = & m\sin\theta \cos\phi ; \\
m^2 & = & m\sin\theta \sin\phi ;\\
m^3 & = & m\cos \theta ;\\
H_1 & = & H \sin\theta_H \cos\phi_H ; \\
H_2 & = & H \sin\theta_H \sin\phi_H ;\\
H_3 & = & H \cos \theta_H .
\ea
\end{subequations}
We use Eqs.~(2.18) in Ref.~\cite{covariant-Langevin-2020} to compute the generalized potential $\tilde U(m ,\theta, \phi )$ and the kinetic matrix $\tilde L_{ab} $ in spherical coordinates: 
\ba
 \tilde U &=& U - \log m^2 \sin \theta, 
\nonumber\\
\mathbf{ \tilde  L}  &=& T
\begin{pmatrix}
0 & 0 & 0\\
0 & \eta & \frac{\gamma_0}{m} \csc\theta\\
0 & - \frac{\gamma_0 \csc\theta} {m} & \eta \csc^2\theta
\end{pmatrix},
\ea
Note that the entries of $\mathbf{ \tilde  L}$ involving $\tilde m^1 = m$  all vanish identically.  This is expected since the dynamics is restricted on the sphere. 


 Using Eqs.~(\ref{FPE-j}) and (\ref{current}), we obtain FPE in spherical coordinates:
\ba
\partial_t p &=& \partial_\theta j^\theta + \partial_\phi j^\phi; \\
j^\theta 
& = & - T \eta \left( \partial_\theta + (\partial_\theta U) \right) p - T \frac{\gamma_0}{m} \csc\theta  (\partial_\phi U)p ; \\
j^\phi 
& = &  -  T \eta \csc^2\theta \left (\partial_\phi +  ( \partial_\phi U ) \right) p + 
T \frac{\gamma_0 \csc\theta} {m}  (\partial_\theta U) p \nonumber\\
&&+ T \frac{\gamma_0}{m} \cos\theta \csc^2\theta p,
\ea
which is in accordance with that given by Brown~\cite{Brown1963}.


\subsection{Detailed balance}
The conditions of detailed balance, which guarantee that a steady state  can be interpreted as a proper thermodynamic equilibrium state, are essential for a proper understanding of non-equilibrium statistical physics.  For covariant Langevin dynamics (\ref{Langevin-def-m}) these conditions were derived in Eqs.~(2.31) of Ref.~\cite{covariant-sto-therm-1}.  In the present case, the magnetization $\mv$ plays the role of state variable, and the magnetic field $\Hv$ plays the role of control parameter.  Both $\mv$ and $\Hv$ are odd under time-reversal.   Hence Eqs.~(2.31) of Ref.~\cite{covariant-sto-therm-1} reduce to  
\begin{subequations}
\label{DB-condition-2}
\ba
\varepsilon_i B^{ij}(-\mv, -\Hv) \varepsilon_j 
&=& B^{ij}(\mv, \Hv ),
\label{DB-condition-B} \\
\varepsilon_i Q^{ij}(-\mv, -\Hv) \varepsilon_j 
&=& - Q^{ij}(\mv, \Hv), \\
 U(-\mv, - \Hv) &=& U(\mv, \Hv), 
 \label{DB-condition-U-1}\\ 
\int_{\mv}e^{-U(\mv, \Hv)} &=&1,
\label{normalization-U}
\ea
\end{subequations}
where $\varepsilon_i  = -1$, and 
\ba
\int_\mv = \int dm^1 dm^2 dm^3 \, \delta (|\mv|^2 - m^2).
\ea
Note that $Q^{ij}, B^{ij}$ are respectively the antisymmetric part and the symmetric part of the kinetic matrix $L^{ij}$. 
 We can explicitly verify that $Q^{ij}, B^{ij}, U$ as given in Eqs.~(\ref{LQBU-LLG}) satisfy the conditions (\ref{DB-condition-2}).  The normalization condition (\ref{normalization-U}) guarantees that $e^{-U}$ can be understood as a physically realizable state.  

As shown in Ref.~\cite{covariant-sto-therm-1},  the conditions (\ref{DB-condition-2}) are equivalent to the condition of microscopic reversibility by Onsager.  For infinitesimal transition time $dt$, the microscopic reversibility conditions can be expressed in terms of transition probabilities: 
\ba
 \log \frac{P_{\Hv} (\mv_1 | \mv; dt) }
{P_{-\Hv}(- \mv | - \mv_1; dt)} 
&=& - U(\mv_1, \Hv ) 
+ U(\mv, \Hv )
\nonumber\\
&=& \Hv \cdot d \mv.
\label{heat-formula-def-0}
\ea  
where $\mv_1 = \mv + d \mv$, and  the subscripts $\Hv$ and $- \Hv$ characterize the magnetic field of the processes. 


\subsection{Stochastic Landau-Lifshitz-Bloch equation}
If the system is not far below the Curie point, the fluctuations of the magnitude of $\mv$ may not be neglected.  In this case, we must modify the dynamics such that the magnitude of $\mv$ can also change. This can be achieved by adding a longitudinal part to the kinetic matrix $B^{ij}$, which controls the dynamics of $|\mv|$.  Hence Eqs.~(\ref{LQBU-LLG-B}) and (\ref{LQBU-LLG-b}) become:
\ba
 B^{ij} &=&  T \eta_\perp \left( m^2  \delta^{ij} - m^i m^j \right)
 + T \eta_\parallel m^i m^j  , \\
b^{i \alpha}&=& 
\frac{\sqrt{2 \eta_\perp T}}{m} (m^2 \delta^{i\alpha}  - m^i m^\alpha)
+ \frac{\sqrt{2 \eta_\parallel T}}{m}  m^i m^\alpha, 
\nonumber\\
\ea
where $\eta_\perp$ and $\eta_\parallel$ are respectively the transverse and longitudinal damping coefficients,  {which are generally different.} 
The asymmetric matrix $Q^{ij}$ remains the same as in Eqs.~(\ref{LQBU-LLG}). 
To penalize the change of $|\mv|$, we also need to add a term $a\, (\mv^2 - \mv_0^2)^2/2$ to the generalized potential.  Hence the new generalized potential becomes
\be
U = - \beta \mv \cdot \Hv + \frac{a}{2}\, (\mv^2 - \mv_0^2)^2 - F (\Hv),
\ee 
where $F (\Hv)$ is fixed by normalization. The resulting soft-spin stochastic differential equation may be called the {\em Landau-Lifshitz-Bloch equation} (sLLB)~\cite{Garanin1997,Evans2012,Atxitia2016}:
\ba
{d \mv} &=&   \gamma_0 T\, (\mv \times {\boldsymbol \nabla} U)  dt
\nonumber\\
&+& \eta_\perp T\, \mv \times (\mv \times  {\boldsymbol \nabla} U)dt
+ \eta_\parallel T\, \mv (\mv \cdot   {\boldsymbol \nabla} U) \, dt
\nonumber\\
&-& 2 T ( \eta_\perp - 2 \eta_\parallel)  \,\mv dt  +  {\sqrt{2 \eta_\perp T}} \, \mv \times d{\boldsymbol W}
\nonumber\\
&+&
 {\sqrt{2 \eta_\parallel T}}\,  \left( \mv \cdot d{\boldsymbol W}  \right) \mv/m.
\label{LLG-stochastic-1}
\ea
 The Fokker-Planck operator associated with this sLLB equation can be similarly constructed.  

\section{Stochastic Thermodynamics}
\label{sec:ST}

In Ref.~\cite{covariant-sto-therm-1} a general formalism of stochastic thermodynamics was developed for the covariant Langevin dynamics (\ref{Langevin-def-m}) which satisfies detailed balance.  We shall apply this formalism  to sLL dynamics described by Eq.~(\ref{Langevin-def-m-vec}).   We recall that the theory in  Ref.~\cite{covariant-sto-therm-1} was established for Langevin dynamics in arbitrary curved space with a metric tensor $g_{ij}$.  In the present case, because of the reason we explained in Sec.~\ref{sec:sLL}, we use the  Cartesian components $m^i$.  Hence the metric tensor should be chosen as the the trivial one $g_{ij} = g^{ij} = \delta_{ij}$.

\subsection{Work, heat and entropy production}

We study non-equilibrium processes during which the magnetic field $\Hv_t$ is systematically varied as a function of time.  {\em The fluctuating internal energy} is defined as:
\be
\mathscr E (\mv, \Hv_t)  \equiv - \mv \cdot \Hv_t, 
\ee
where $ \Hv_t$ is the instantaneous magnetic field.  The definition of energy is such that at any time $t$, the instantaneous equilibrium distribution (\ref{p-eq-U-1}) takes the usual form of Gibbs-Boltzamnn distribution:
\ba
p^{\rm eq}(\mv; \Hv_t)  
&=& e^{- \beta  \mathscr E(\mv, \Hv_t) + \beta F(\Hv_t) }.
\label{p_EQ-2}
\ea

{\em The non-equilibrium internal energy} is defined as the ensemble average of $\mathscr E (\mv, \Hv_t) $:
\ba
E[p] &\equiv& - \int_\mv (\mv \cdot \Hv) p(\mv, t). 
\label{TM-state-functions-def-E}
\ea
{\em The non-equilibrium entropy} is defined as:
\ba
S[p] &\equiv& - \int_{\mv} p  (\mv,t)  \, \log p  (\mv,t). 
\label{TM-state-functions-def-S}
\ea
{\em The non-equilibrium free energy} is defined as
\ba
F[p]  &\equiv&  E[p] - T S[p] 
\nonumber \\
&=& \int_{\mv} p  (\mv,t)
 \left( - \mv \cdot \Hv  + T  \log p  (\mv,t)
\right). 
\label{TM-state-functions-def-F}
\ea 
For equilibrium state Eq.~(\ref{p_EQ-2}), $F[p]$ reduces to  equilibrium free energy $F(\Hv)$, which is given by Eq.~(\ref{F-lambda-def}). 

 {\em The differential work and heat at the trajectory level} are defined as 
\ba
\dbar \mathscr W &\equiv& d_{\Hv} \mathscr E 
 = - \mv \circ d \Hv,
\label{dW-traj-def-1}\\
\dbar \mathscr Q &\equiv& d_{\mv} \mathscr E  = - \Hv \circ d \mv,
\label{dQ-traj-def-1}
\ea
where $d_{\Hv}, d_{\mv}$ mean respectively differentials with respect to $\Hv$  and $\mv$~\cite{covariant-sto-therm-1}, and $\circ$ means inner product in the Stratonovich's sense. Since $\Hv$ is independent of the magnetization $\mv$, the Stratonovich product $\circ$ can be replaced by the usual scalar product $\cdot$ .   We have the first law of thermodynamics at the trajectory level:
\ba
d\mathscr E  = \dbar \mathscr W + \dbar\mathscr Q . 
\label{1st-law-traj-d}
\ea
Furthermore, using sLL equation (\ref{Langevin-def-m-vec}), we may rewrite the differential heat (\ref{dQ-traj-def-1}) as
\ba
\dbar \mathscr Q &=& \Hv \cdot \big[ \eta \, 
\mv \times (\mv \times \Hv)\, dt 
\nonumber\\
&+&  2 T \eta \,\mv \,dt  
-  {\sqrt{2 \eta T}} \mv \times d{\boldsymbol W} \big], 
\ea
which may be understood as the work done by the heat bath.  This is of course consistent with the common understanding of heat in stochastic thermodynamics. In the limit $\eta \rightarrow 0$, the system is completely decoupled from the heat bath, and the heat vanishes identically.  
 
We can integrate the first law Eq.~(\ref{1st-law-traj-d}) along a trajectory $\bm \gamma$, and obtain:
\ba
\Delta \mathscr E = \mathscr W [\bm \gamma] 
+ \mathscr Q [\bm \gamma],
 \label{1st-law-int}
\ea
where $\Delta \mathscr E $  as the total change of the energy along $\bm \gamma$: 
\ba
\Delta \mathscr E  \equiv \mathscr E(\mv(\tau), \Hv_\tau)
- \mathscr E(\mv(0), \Hv_0), 
\label{Delta-E-def}
\ea
whereas $ \mathscr W [\bm \gamma] $ and $\mathscr Q [\bm \gamma]$ are the integrated work and heat:
\ba
 \mathscr W [\bm \gamma] &\equiv& 
   \int_{\gamma} \dbar \mathscr W
 = - \int_\gamma \mv(t) \cdot d \Hv_t,
 \label{W-path-def}\\
  \mathscr Q [\bm \gamma] &\equiv& 
   \int_{\gamma} \dbar \mathscr Q
=  { - \int_\gamma  d \Hv_t \cdot \mv(t) . }
 \label{Q-path-def}
\ea
Equation (\ref{1st-law-int}) is the first law for a finite trajectory. 

Using Eq.~(\ref{dQ-traj-def-1}), we can rewrite Eq.~(\ref{heat-formula-def-0}) as
\ba
 \log \frac{P_{\Hv} (\mv_1 | \mv; dt) }
{P_{-\Hv}(- \mv | - \mv_1; dt)} 
=  - \beta \dbar \mathscr Q = d  S^{\rm env},
\label{heat-formula-def-1}
\ea 
where $d  S^{\rm env}$ is the differential of environmental entropy during the infinitesimal step of evolution.   The second equality follows because the environment is a heat bath always in equilibrium.  We shall call Eq.~(\ref{heat-formula-def-1}) {\em the condition of local detailed balance}.

 Work and heat at the ensemble level are defined as the ensemble averages of work and heat at the trajectory level.  Here ensemble average means averaging both over probability distribution of $\mv$ and, if necessary, over noise at a given time.  These quantities can be written as
\begin{subequations}
\ba
\dbar Q &=& \langle \!\langle \dbar \mathscr Q \rangle \!\rangle
= \int_{\mv} \mathscr E   \LFP   p \, dt,
\label{dQ-def-ensemble} \\
\dbar W &=& \langle \!\langle \dbar \mathscr W \rangle \!\rangle
= - \int_{\mv} p \, \mv \cdot d \Hv, \quad\quad
\ea
\label{dQ-dW-def-ensemble}
\end{subequations}
\!\!\!\!where $\LFP  $ is the Fokker-Planck operator (\ref{FPO-LLG}). We then have the first law at the ensemble level is:
\be
d E = \dbar Q + \dbar W .
\label{dE}
\ee

Taking the differential of Eq.~(\ref{TM-state-functions-def-F}), we find 
\ba
dF  = \dbar W + \dbar Q - T d S. \label{dF}
\ea
The change of the joint entropy of the system and bath, i.e. {\em the entropy production}, at the ensemble level, is 
\ba
dS^{\rm tot} =  d S  - \beta \dbar Q 
=  \beta (\dbar W  - d F ).  
\label{dS^tot-1-XY}
\ea
Invoking Eq.~(3.11) of  Ref.~\cite{covariant-sto-therm-1}, the rate of entropy production  is given by
\ba
\Sigma^{\rm tot}  &\equiv&  \frac{dS^{\rm tot} }{dt}
 \nonumber \\
 &=& \int_\mv \left[ (\partial_i + \partial_i U)p\right]
   \frac{B^{ij}}{p}
\left[ (\partial_j + \partial_j U)p \right],
  \label{dS^tot-1}
\ea
which, up on using Eqs.~(\ref{LQBU-LLG}) can be rewritten into
\ba
\Sigma^{\rm tot}  &=& T \eta \Big\langle m^2 \big( 
\boldsymbol \nabla \log p - \beta \Hv \big)^2
\nonumber\\
& -& \big( \mv \cdot \big( 
\boldsymbol \nabla \log p 
- \beta \Hv \big) \big)^2 \Big\rangle.
 \label{dS^tot-2}
\ea
As one can easily see, $\Sigma^{\rm tot}$ is non-negative, and  vanishes identically only at equilibrium state (\ref{p_EQ-2}).  

Invoking Eqs.~(\ref{j_R-IR-def}), we may also rewrite the entropy production rate (\ref{dS^tot-1}) as
\ba
\Sigma^{\rm tot}  =  \int_\mv {p}^{-1} j^i_{\rm IR}
   {B_{ij}} j^j_{\rm IR}, 
  \label{dS^tot-3}
\ea
where $B_{ij}$ is the generalized inverse of $B^{ij}$.   This explicitly demonstrates that the entropy production depends only on the irreversible current $ j^j_{\rm IR}$ but not on the reversible current $ j^j_{\rm R}$.   In the equilibrium, the irreversible current vanishes identically, hence also does the entropy production. 

 {Our Eqs.~(\ref{dS^tot-1})-(\ref{dS^tot-3}) should be contrasted with the corresponding result in Ref.~\cite{Bando2015}).  The entropy production postulated in Ref.~\cite{Bando2015} (the unlabeled equation after Eq.~(8)) is a quadratic function of the probability current, which is non-vanishing even at thermal equilibrium.  This is of course thermodynamically inconsistent.  }

\subsection{Fluctuation Theorem}
\label{sec:FT}
Consider a {\em forward process} where the system starts at $t = 0$ from the initial equilibrium state $p^{\rm eq}(\mv; \Hv_0)$ (defined in Eq.~(\ref{p_EQ-2})), and the field evolves according to the {\em forward protocol} $\Hv_t$, until $t = \tau$, when the process stops.  We define the {\em backward process} such that the system starts at $t = 0$ from the initial equilibrium state $p^{\rm eq}(\mv; - \Hv_\tau)$ and the field evolves according to the {\em backward protocol} $\Hv^{\rm B}_t \equiv - \Hv_{\tau - t}$, until $t = \tau$, when the process stops.   Note that both the forward process and the backward process take place in the time interval $[0, \tau]$.  In general, the system is not in equilibrium either at the end of the process or at the end of the backward process. 

Consider a {\em forward trajectory}:
\ba
 \bm \gamma &=& \{\mv(t), \,\, t \in [0, \tau] \},
 \ea
 we define  its  {\em backward trajectory} as
 \ba
\hat {\bm \gamma} &=&  \{ - \mv(\tau - t),\,\,  t \in [0, \tau] \}. 
\ea 
Let  $\mathscr W_{\rm F} [ \bm \gamma], \mathscr Q_{\rm F} [ \bm \gamma]$ ($\mathscr W_{\rm B} [\hat {\bm \gamma}],\mathscr Q_{\rm B} [\hat {\bm \gamma}] $) be the integrated work and heat along $\bm\gamma$ ($\hat {\bm \gamma}$) in the forward (backward) process, which can be readily obtained by integrating the differential work and heat that are defined in Eqs.~(\ref{dW-traj-def-1}) and (\ref{dQ-traj-def-1}).  We easily find the following symmetry:
\ba
\mathscr W_{\rm F} [ \bm \gamma] = 
- \mathscr W_{\rm B} [\hat {\bm \gamma}] 
=  - \int_\gamma \mv \circ d \Hv,
\label{W-F-decomp-1}\\
\mathscr Q_{\rm F} [ \bm \gamma] = 
- \mathscr Q_{\rm B} [\hat {\bm \gamma}] 
=  - \int_\gamma \Hv  \circ d\mv . 
\label{Q-F-decomp-1}
\ea

We further introduce ${\bm\gamma}_0  \equiv \mv(0)$ and $\hat {\bm\gamma}_0 \equiv - \mv(\tau)$ to denote the initial state of $ \bm\gamma, \hat {\bm \gamma}$, respectively.  These notations (boldface) should be carefully distinguished from $\gamma_0$, the {\em gyromagnetic ratio}, appearing in Eqs.~(\ref{LLG-0}) and (\ref{Langevin-def-m-vec}).  We can construct the pdfs of trajectories both for the forward process and for the backward process,  using the definition of conditional probability:
\begin{subequations}
 \label{p_FB-gamma}
\ba
p_{\rm F}[ \bm \gamma] &=& 
p_{\rm F}[\bm \gamma | {\bm\gamma}_0] \,
 p^{\rm eq}(\mv(0); \Hv_0),
 \label{p_F-gamma}\\
 p_{\rm B}[\hat {\bm \gamma}] &=& 
p_{\rm B}[\hat {\bm \gamma} |\hat  {\bm\gamma}_0]  \,
p^{\rm eq}(- \mv(\tau); - \Hv_\tau),
 \label{p_B-gamma}
\ea
\end{subequations}
where $p_{\rm F}[\bm \gamma | {\bm\gamma}_0], p_{\rm B}[\hat {\bm \gamma} |\hat  {\bm\gamma}_0]$ are the conditional pdf of trajectories of the forward (backward) processes given their initial states, whilst $ p_{\rm B}$ is defined in Eq.~(\ref{p_EQ-2}). 

Because of the Markov property, $p_{\rm F}[\bm \gamma  |{\bm\gamma}_0]$ and $ p_{\rm B}[\hat {\bm \gamma} |\hat  {\bm\gamma}_0] $ can be calculated using the time-slicing method.   Further using Eq.~(\ref{heat-formula-def-1}) for each pair of time-slices, we have
\ba
\log  \frac{ p_{\rm F}[\bm \gamma | {\bm\gamma}_0] }  
{ p_{\rm B}[\hat {\bm \gamma} |\hat  {\bm\gamma}_0] } 
= {- \beta \mathscr Q_{\rm F}[ \bm \gamma]}
= \int_\gamma d S^{\rm env}, 
\label{p-gamma-cond-ratio}
\ea
where $\mathscr Q_{\rm F}[ \bm \gamma]$ is the total heat absorbed by the system along the trajectory $\bm \gamma$ in the forward process. 

Let us define: 
\ba
\Sigma_{\rm F}[ \bm \gamma] &\equiv&
  \log \frac{p_{\rm F}[ \bm \gamma]}{p_{\rm B}[\hat {\bm \gamma}]} .
  \ea
Using Eqs.~(\ref{p_FB-gamma}) and (\ref{p-gamma-cond-ratio}), we obtain:
  \ba
\Sigma_{\rm F}[ \bm \gamma] 
&=&  \log \frac{ p^{\rm eq}(\mv(0); \Hv_0)}
{p^{\rm eq}(- \mv(\tau); - \Hv_\tau)}
-  \beta \mathscr Q_{\rm F}[ \bm \gamma]
\quad \label{p-gamma-ratio-1} 
\ea

Recalling the symmetry: ${p^{\rm eq}(- \mv; - \Hv)}
= {p^{\rm eq}( \mv;  \Hv)}$, if the protocol is such that the final state  of the forward process is the {equilibrium state} corresponding to $\Hv_\tau$, we may also write Eq.~(\ref{p-gamma-ratio-1}) into 
\ba
\Sigma_{\rm F}[ \bm \gamma] = 
- \log \frac{ p(\mv(\tau), \tau)}{p(\mv(0), 0)}
- \beta \mathscr Q_{\rm F} [ \bm \gamma],
\label{Sigma-gamma-1}
\ea
which is {\em the stochastic entropy production}~\cite{Seifert-2005} along the trajectory $\bm \gamma$ in the forward process.  If the system is not in the NESS at the end of the forward process, however, the physical meaning of $\Sigma_{\rm F}[ \bm \gamma] $ is more subtle. 

Further taking advantage of Eq.~(\ref{p_EQ-2}) as well as the first law (\ref{1st-law-int}), we may rewrite Eq.~(\ref{p-gamma-ratio-1}) into:
\ba
\Sigma_{\rm F}[ \bm \gamma] = 
\log \frac{p_{\rm F}[ \bm \gamma]}{p_{\rm B}[\hat {\bm \gamma}]} 
= \beta\left( \mathscr W_{\rm F} [\gamma]  - \Delta F \right), 
\label{p-gamma-ratio-1-1}
\ea
where $\Delta F$ is defined as
\ba
\Delta F \equiv F (\Hv_\tau)  - F(\Hv_0). \label{Delta-F-def}
\ea

\begin{figure}[t!]
    \centering
    \includegraphics[width=3.4in]{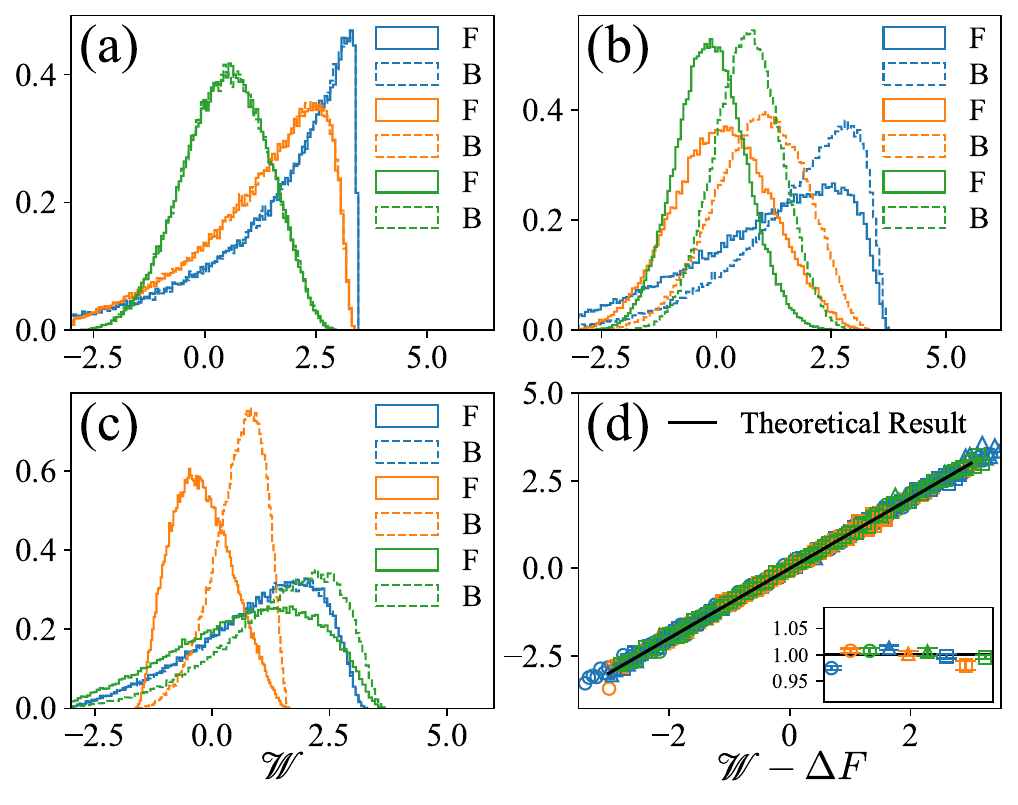}
    \caption{Verification of FT (\ref{CFT-eq}) for the integrated work . (a), (b), (c): Histograms of the work $\mathscr{W}$, where all processes are defined in Table \ref{tab::ft/ft_f_combined_TR}.      In all legends F, B mean forward and backward respectively.  (d): Verification of FT (\ref{CFT-eq}), where the vertical axis is $\log p_{\rm F} (\mathscr{W})/p_{\rm B}(-\mathscr{W})$. The black straight-line is the FT (\ref{CFT-eq}). 
    Circles, triangles, and squares are respectively data from panels (a), (b), (c).  Inset: The fitting slopes and error bars for each process.}
    \label{fig::ft/ft_f_combined_TR}
\end{figure}

Using Eq.~(\ref{W-F-decomp-1}), we may rewrite Eq.~(\ref{p-gamma-ratio-1-1}) into
\ba
\log \frac{p_{\rm F}[ \bm \gamma]}{p_{\rm B}[\hat {\bm \gamma}]} 
&=&  \beta  \left( \mathscr W_{\rm F} [ \bm \gamma] - \Delta F \right)
\nonumber\\
&=&  - \beta \left(  \mathscr W_{\rm B} [\hat {\bm \gamma}]
+  \Delta F \right).  \quad\quad
\label{DFT-1}
\ea

We can now define the pdfs of the integrated work both for the forward process and for the backward process:
\ba
p_{\rm F}( \mathscr W) &\equiv& \int D {\bm \gamma} \, 
\delta \left( \mathscr W - \mathscr W_{\rm F}[\bm \gamma] \right) \, p_{\rm F}[\bm \gamma], \\
p_{\rm B}( \mathscr W) &\equiv& \int D {\bm \gamma} \, 
\delta \left( \mathscr W - \mathscr W_{\rm B}[\bm \gamma] \right) \, p_{\rm B}[\bm \gamma]. 
\ea
Taking advantage of Eq.~(\ref{DFT-1}), and using standard methods of stochastic thermodynamics, we can prove the following relation for $\mathscr W$:
\ba
p_{\rm F}(\mathscr W) e^{ - \beta \mathscr W 
+ \beta \Delta F}
= p_{\rm B}(-\mathscr W) , 
\label{CFT-eq}
\ea
where $\Delta F$ is defined in Eq.~(\ref{Delta-F-def}).   This is {\em Crooks Fluctuation Theorem}.  Integrating Eq.~(\ref{CFT-eq}) over $\mathscr W$, we derive {\em Jarzynski's equality}: 
\be
\left\langle e^{-\beta \mathscr W} \right\rangle 
= e^{-\beta \Delta F },
\label{JE}
\ee
where the average in the LHS is taken with respect to the forward process.  

\begin{table}[t!]
    \centering
    \renewcommand{\arraystretch}{1.4}
    \tabcolsep=7pt
    \begin{tabular}{|c|c|c|c|c|c|}
     \hline
    process & color & $\eta$ & $\tau$ & {$\boldsymbol{H}_\tau$} \\ \hline
    \multirow{3}{*}{(a)} & blue & 0.05 & \multirow{3}{*}{1} &  \multirow{3}{*}{(0,\,0,\,-1)} \\ \cline{2-3}
    & orange & 0.5 &  &  \\ \cline{2-3}
    & green & 3 &  &  \\ \hline
    \multirow{3}{*}{(b)} & blue & \multirow{3}{*}{0.5} & 1 &  \multirow{3}{*}{(0,\,-1,\,-1)} \\ \cline{2-2} \cline{4-4}
    & orange &  & 5 &  \\ \cline{2-2} \cline{4-4}
    & green &  & 10 &  \\ \hline
    \multirow{3}{*}{(c)} & blue & \multirow{3}{*}{1} & \multirow{3}{*}{1} &  (0,\,0,\,-1) \\ \cline{2-2} \cline{5-5} 
    & orange & &  & (0,\,1,\,1) \\ \cline{2-2} \cline{5-5} 
    & green & &  & (0,\,-1,\,-1) \\ \hline
    \end{tabular}
         \vspace{3mm}
   \caption{ Protocols for verifying FT (\ref{CFT-eq}).  In all processes, the field starts from $\boldsymbol{H}_0 = (0,\,0,\,1)$ and linearly increased up to $\boldsymbol{H}_\tau$.  In all smulations $T =1$.}
    \label{tab::ft/ft_f_combined_TR}%
\end{table}%

We numerically simulate the sLL equation to verify the fluctuation theorem (\ref{CFT-eq}). The details of the simulation is given in App.~\ref{sec:numerical}. We simulated three non-equilibrium processes and six dynamic protocols as shown in Table ~\ref{tab::ft/ft_f_combined_TR}.    {In all these protocols, the magnetic field changes linearly with time:
\ba
\Hv_t = \Hv_0 + \frac{t}{\tau} \left( \Hv_\tau  - \Hv_0\right). 
\ea  }  
Distributions of the integrated works both for the forward process and for the backward process are shown in Fig.~\ref{fig::ft/ft_f_combined_TR}.  The pdfs of the integrated work are shown in Fig.~\ref{fig::ft/ft_f_combined_TR} (a) (b) and (c), whereas the fluctuation theorem Eq.~(\ref{CFT-eq}) is verified in Fig.~\ref{fig::ft/ft_f_combined_TR} (d).  
As one can see, the agreement between simulation and theory is remarkable.  

\section{Conclusion}
\label{sec:conclusion}

In this work, we have developed a theory of stochastic thermodynamics for micromagnetism, whose dynamics is described by sLL equation.  This theory is a direct application of the covariant stochastic thermodynamics developed in Ref.~\cite{covariant-Langevin-2020,covariant-sto-therm-1}, with the peculiar features that both dynamics variables and control parameters are odd under time-reversal, and with the noises being multiplicative.   Nonetheless, the dynamics satisfies detailed balance, and is able to achieve thermal equilibrium at fixed magnetic field. 

In the next work, we shall study stochastic thermodynamics for micro-magnet driven by spin torque.  Such a system lacks detailed balance, which makes the theory of stochastic thermodynamics more complex and more interesting. 

The authors acknowledge support from NSFC via grant \#12375035(X.X.), as well as Shanghai Municipal Science and Technology Major Project (Grant No.2019SHZDZX01).  

\appendix
\section*{Details of numerical simulations}
\label{sec:numerical}
A detailed discussion of numerical simulations of micromagnetics can be found in chapter 11 of Ref.~\cite{Mayergoyz-book-2009}. Here use the first-order Euler-Maruyama scheme \cite{Kloeden1992} to numerically solve  Eq.(\ref{LLG-0}). The time step is chosen as $\Delta t= 10^{-3}$, $\gamma_0 = 1$, $T=1$, and $\eta$ can take on different values. The magnitude of $\mv$ is fixed at $|\boldsymbol{m}|=\sqrt{3}$.

The discretized dynamical equation is:
\begin{align}
    \Delta \boldsymbol{m}(t_{n+1}) =&
     - \gamma_0\, \boldsymbol{m}(t_{n})\times
       \boldsymbol{H}(t_{n}) \Delta t 
       \\
    & - \eta \, \boldsymbol{m}(t_{n})\times
    [\boldsymbol{m}(t_{n}) \times 
    \boldsymbol{H}(t_{n})]  \Delta t \\
    & -2T \, \eta \,
    \boldsymbol{m}(t_{n})\Delta t
    + \sqrt{2\eta T \Delta t}\boldsymbol{m}(t_{n})
     \times \boldsymbol{\xi} .
\end{align}
where $\boldsymbol{\xi}=(\xi_1,\xi_2,\xi_3)$  are normalized Gaussian random numbers generated by the mt19937 algorithm~\cite{matsumoto1998mersenne}.

After each step, $\boldsymbol{m}(t_{n})$ is normalized to the initial $|\boldsymbol{m}(0)|$
\begin{equation}
    \boldsymbol{m}(t_{n + 1})= \frac{\boldsymbol{m}(t_{n})+\Delta \boldsymbol{m}(t_{n}+\Delta t)}{|\boldsymbol{m}(t_{n})+\Delta \boldsymbol{m}(t_{n}+\Delta t)|} |\boldsymbol{m}(0)| .
\end{equation}

Numerically, we calculate the work in Eq.~(\ref{W-path-def}) using a discretized formula as
\begin{equation}
    \mathscr{W}= \sum_{i=0}^{N-1} - \frac{\boldsymbol{m}(t_{i+1})+\boldsymbol{m}(t_{i})}{2}\cdot (\boldsymbol{H}(t_{i+1})-\boldsymbol{H}(t_i)),
\end{equation}
where the process starts with $t_0=0$ and ends at $t_N = \tau$.

\end{document}